\documentclass[
	a4paper, 
	10pt, 
	unnumberedsections, 
	twoside, 
]{LTJournalArticle}

\runninghead{} 

\usepackage{tikz}

\title{EMiX: Emulating Beyond Single-FPGA Limits}

\author{%
Alexander Kropotov, 
Miquel Moreto,
Behzad Salami
}

\date{BSC\\ 
}

\renewcommand{\maketitlehookd}{%
	\begin{abstract}
		\noindent
FPGA-level emulation is a key step in pre-silicon chip design validation. However, emulating large-scale multi-core systems increasingly exceed the hardware resource capacity of a single FPGA, limiting the feasibility of full-system emulation. To address this challenge, we introduce EMiX, a scalable multi-FPGA framework that enables distributed emulation of multi-core RISC-V architectures beyond single-FPGA resource limits. EMiX systematically partitions a monolithic multi-core design into multiple components and deploys them across multiple interconnected FPGAs, effectively exploiting inter-FPGA interconnects to balance scalability and performance without requiring fundamental RTL redesign. We prototype EMiX with a 64-core architecture across eight interconnected Alveo U55c FPGAs (scalable on core and FPGA counts), successfully demonstrating full-system execution including Linux boot. EMiX will be released as an open-source platform.
\end{abstract}
}

\begin{document}

\maketitle 


\section{Motivation}
Emulating modern multi-core systems increasingly exceed the resource capacity of a single FPGA. Scaling such systems requires distributing the design across multiple FPGAs, introducing challenges in design partitioning, inter-FPGA communication, and preserving software stack functionality and visibility. Prior academic frameworks such as SMAPPIC~\cite{chirkov2023smappic}, FireAxe~\cite{whangbo2024fireaxe}, and HeteroProto~\cite{zhang2025heteroproto} explore multi-FPGA prototyping but often rely on limited inter-FPGA interconnects, e.g., PCIe or proprietary infrastructures, e.g., Amazon EC2, which can restrict scalability, performance, and accessibility. Industrial platforms such as Synopsys HAPS, Siemens Veloce, and Cadence Palladium provide powerful environments supporting automated input RTL partitioning, efficient inter-FPGA interconnection but are not open source and accessible for research. EMiX aims to bridge this gap by introducing an efficient multi-FPGA emulation platform for many-core systems with unique features: \textbf{\textit{i)}} leveraging heterogeneous inter-FPGA interconnects to enable multi-core emulation, balancing performance and scalability, \textbf{\textit{ii)}} supporting full-system OS and software stack, and \textbf{\textit{iii)}} open-source and developed on commodity FPGA hardware infrastructure, successfully tested with 64-cores 8-FPGAs, parametrizable to many-core many-FPGAs.

\section{EMiX Architecture}

\begin{figure*}[htbp]
  \centering
  \includegraphics[width=2.0\columnwidth]{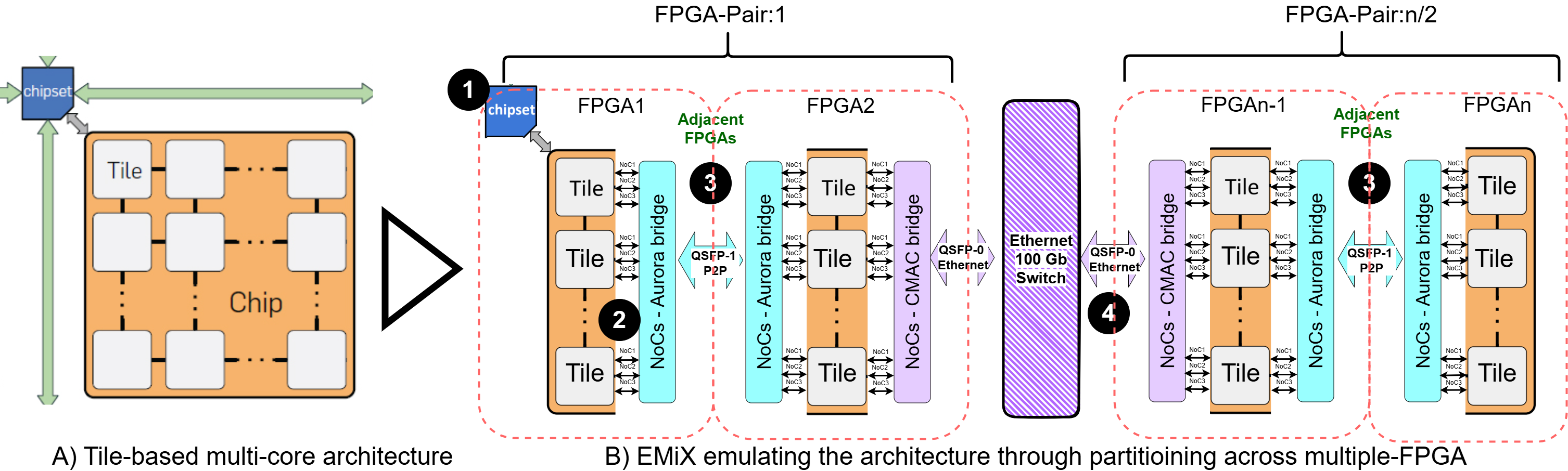}
  
\caption{EMiX multi-core multi-FPGA emulation (in this figure illustared for Vertical partitioning), prototyped on Makinote hardware infrastructure~\cite{perdomo2024makinote}: 
\textcircled{1} Chipset interface to access the emulated system at OS level; 
\textcircled{2} NoC partition boundaries; 
\textcircled{3} Low-latency Aurora P2P links over QSFP-1 via the NoC-Aurora bridge for adjacent FPGAs; 
\textcircled{4} Scalable Ethernet links over QSFP-0 via the NoC-CMAC bridge for cross-connect across all FPGAs interconnect using 100Gb Ethernet switch.}

  \label{fig:emix}
\end{figure*}

EMiX partitions monolithic tiled many-core RTL designs across multiple interconnected FPGAs. The initial prototype targets OpenPiton~\cite{balkind2016openpiton}, a state-of-the-art open-source many-core platform in which tiles are interconnected through three on-chip networks (NoCs) arranged in a 2D mesh topology. Each tile integrates a RISC-V core (an in-house core in the current EMiX prototype), L1/L2 caches, and NoC routers, and the architecture can scale up to 256 tiles per dimension. The NoCs use 64-bit unidirectional links with wormhole routing, while a chip bridge connects the tile array to off-chip peripherals such as UART, memory, and Ethernet to enable software interaction. EMiX partitions the system along tile boundaries at NoC edges and distributes tiles across multiple FPGAs. Each FPGA can host one or more tiles, and the partitioning strategy (e.g., horizontal or vertical) is configurable. 

The current prototype of EMiX is deployed on Makinote~\cite{perdomo2024makinote}, a large-scale FPGA cluster composed of 96 Alveo U55C devices, which are equipped with two QSFP interfaces. In Makinote, the first interface (QSFP-0) connects all FPGAs through 100 Gb Ethernet switches, enabling full network connectivity across the cluster. The second interface (QSFP-1) provides direct point-to-point (P2P) connections between adjacent FPGAs. As illustrated in Figure~\ref{fig:emix}, EMiX exploits both interfaces to combine scalable connectivity (through QSFP-0) with low-latency communication (through QSFP-1). In practice, EMiX organizes FPGA communication into two levels:

\begin{itemize}

\item Low-Latency Path (Aurora-based direct P2P):
At the first level, pairs of adjacent FPGAs are interconnected through the low-latency QSFP-1 links. These links use AMD’s Aurora IP\footnote{https://www.amd.com/en/products/adaptive-socs-and-fpgas/intellectual-property/aurora64b66b.html}
, which implements a lightweight protocol over GTY SerDes transceivers and communicates directly via optical cables. We develop NoC–Aurora bridge to transfer NoC packets over the Aurora links.

\item High-Scalable Path (Ethernet-based cross-connected):
 At the second level, FPGA pairs communicate through 100 Gb Ethernet links over the QSFP-0 interface, enabling full cross-connectivity across the cluster. EMiX employs AMD’s 100 Gb CMAC IP\footnote{https://www.amd.com/en/products/adaptive-socs-and-fpgas/intellectual-property/cmac\_usplus.html}. EMiX is equipped with NoC–CMAC bridge to transports NoC packets across the Ethernet infrastructure.

\end{itemize}

By exploiting both QSFP interfaces of Alveo U55c FPGAs, EMiX maintains full connectivity among FPGA nodes, providing scalability for many-core many-FPGAs emulation. Also, this dual-channel design provides more benefits: at run-time it reduces the traffic on Ethernet network and provides lower latency for data transfer, and at compile time it simplifies the design implementation by reducing congestion in CMAC-based channels. Finally, the FPGA hosting the chipset uses the Ethernet interface and preserves the software-level access to the emulated platform.

Both communication paths operate on a unified AXI-Stream transport abstraction, which serves as the internal data transport layer within EMiX. The AMD AXI-Stream infrastructure is used for multiplexing, channel mapping, clock-domain crossing (CDC), and demultiplexing of NoC channels participating in cross-FPGA partitioning. Conversion bridges, i.e., NOC-CMAC and NOC-Aurora translate OpenPiton NoC packets to and from Ethernet or Aurora frames. In the Ethernet case, frames include FPGA-specific source and destination MAC addresses, and reliable delivery is ensured through standard Ethernet mechanisms with retransmission of lost frames when necessary.

\section{Experimental Analysis}
The current EMiX prototype emulates a 64-core configuration across 8 FPGAs (4 pairs of Aurora-connected FPGAs interconnected over Ethernet, with 8 cores per FPGA). Both the core count and FPGA count are configurable. In the partitioned system, the first FPGA hosts all required peripherals, including UART, HBM, and Ethernet. A bare-metal multi-core application detects all initialized cores and sequentially runs a memory test on each of them. The prototype is also capable of booting Linux in approximately 15 minutes (vs. 5 minutes for single-FPGA scenario) at a clock frequency of 50 MHz. Under Linux, the integrated 100 Gb Ethernet interface operates correctly, supporting standard networking utilities such as ping and scp, used for software level users accessing to the system. For this prototype, the LUT utilization rate of each FPGA reaches approximately 55\% with an overhead of approximately 16\% mainly used by CMAC, Aurora, NOC-CMAC and NOC-Aurora IPs. 

\section{Discussions}
The current prototype of EMiX introduces several  features including i) multi-core emulation on multi-FPGA platform with flexibility on numbers of tiles, number of FPGAs, and type of design partitioning vertical/horizontal, ii) dual-channel inter-FPGA communication that optimizes scalability and performance, and iii) software-accessible emulation platform capable of supporting OS and full software stack. Future work will extend EMiX with additional capabilities such as a comprehensive performance evaluation of scalability and performance overheads.

\section{Acknowledgments}
This work has been co-financed by the Barcelona Zettascale Laboratory under project reference REGAGE22e00058408992, with support from the Spanish Ministry for Digital Transformation and Public Services, within the framework of the Recovery and Resilience Facility and the European Union – NextGenerationEU; from the research projects ST4HPC (PID2023-147979NB-C21) and the ACAP project (PID2023-146511NB-I00) sup- ported by the Spanish Ministry of Science and Innovation (MICIU), the State Research Agency (AEI), and the European Regional De- velopment Fund (FEDER).

\printbibliography 

\end{document}